\documentclass[conference]{IEEEtran}
\IEEEoverridecommandlockouts
\usepackage{cite}
\usepackage{amsmath,amssymb,amsfonts}
\usepackage{algorithmic}
\usepackage{graphicx, caption, subcaption}
\usepackage{textcomp}
\usepackage{xcolor}
\usepackage{hyperref}
\usepackage[shortlabels]{enumitem}
\def\BibTeX{{\rm B\kern-.05em{\sc i\kern-.025em b}\kern-.08em
    T\kern-.1667em\lower.7ex\hbox{E}\kern-.125emX}}
\begin{document}

\title{Investigating Perceived and Mechanical Challenge in Games Through Cognitive Activity}

\author{\IEEEauthorblockN{1\textsuperscript{st} Christine Hegedues}
\IEEEauthorblockA{\textit{IT-University of Copenhagen}\\
Copenhagen, Denmark \\
hege@itu.dk}
\and
\IEEEauthorblockN{2\textsuperscript{nd} Joao Pedro Dias Constantino}
\IEEEauthorblockA{\textit{IT-University of Copenhagen}\\
Copenhagen, Denmark \\
joao@itu.dk}
\and
\IEEEauthorblockN{3\textsuperscript{rd} Laurits Dixen}
\IEEEauthorblockA{\textit{Center for Digital Play} \\
\textit{IT-University of Copenhagen}\\
Copenhagen, Denmark \\
ldix@itu.dk}
\and
\IEEEauthorblockN{4\textsuperscript{th} Paolo Burelli}
\IEEEauthorblockA{\textit{Center for Digital Play} \\
\textit{IT-University of Copenhagen}\\
Copenhagen, Denmark \\
pabu@itu.dk}
}

%\IEEEoverridecommandlockouts
%\IEEEpubid{\makebox[\columnwidth]{979-8-3503-2277-4/23/\$31.00~\copyright2023 IEEE \hfill}
%\hspace{\columnsep}\makebox[\columnwidth]{ }}

\maketitle

%\IEEEpubidadjcol

\begin{abstract}
Game difficulty is a crucial aspect of game design, that can be directly influenced by tweaking game mechanics. Perceived difficulty can however also be influenced by simply altering the graphics to something more threatening. Here, we present a study with 12 participants playing 4 different minigames with either altered graphics or mechanics to make the game more difficult. Using EEG bandpower analysis, we find that frontal lobe activity is heightened in all 4 of the mechanically challenging versions and 2/4 of the visually altered versions, all differences that do not emerge from the self-reported player experience. This suggests that EEG could aid researchers with a more sensitive tool for investigating challenge in games.

\end{abstract}


\section{Introduction}

Challenge is a crucial element in creating an engaging and immersive gaming experience. 
It serves as a core component of the flow state, a psychological concept introduced by Csikszentmihalyi~\cite{csikszentmihalyi_beyond_1975}, which describes a state of complete absorption in an activity, leading to an optimal experience. While game challenges are typically associated with mechanical aspects such as gameplay mechanics, difficulty settings, and level design, they can also be manifested through visual elements. Visual cues that evoke feelings of danger or threat are often incorporated into more difficult levels, subtly increasing the perceived challenge and overall game experience~\cite{burelli_virtual_2013}. Despite the widespread use of visual elements to enhance difficulty in games, there is a lack of empirical research examining the relationship between visual and mechanical difficulty. Investigating the differences and similarities between these two aspects of challenge can provide valuable insights for game designers and inform us in our understanding of the cognitive and emotional processes underlying gameplay experiences. To evaluate the difficulty levels in games, researchers and designers often employ various surveys, such as the Game Experience Questionnaire (GEQ; \cite{law_systematic_2018}) or the Challenge Originating from Recent Gameplay Interaction Scale (CORGIS \cite{denisova_measuring_2020}). These questionnaires provide insights into players' subjective experiences and their perceptions of the gameplay.

However, surveys have limitations. Firstly, administering surveys can be time-consuming, disrupting the flow of the gameplay experience. Secondly, they measure a retrospective recollection of the experience, which is summative and potentially inaccurate. Survey responses are inherently subjective, making it difficult to compare and generalize results across different contexts and individuals. Additionally, surveys may not capture the full scope of the cognitive and emotional processes that occur during gameplay, which could be better understood using psychophysiological measures. Electroencephalography (EEG) bandpower analysis offers a method for measuring the cognitive state of the player during gameplay. Alpha waves (8-12 Hz) are often linked to idleness or inactivity (for example at shut eyes), reflecting a state of relaxation or disengagement. In contrast, gamma waves (30-60 Hz) are associated with increased engagement and cognitive processing in the corresponding brain areas \cite{abhang_chapter_2016}. Thus, we would expect that playing a more challenging version of a game would result in a higher gamma activity and lower alpha activity, indicating increased attention or engagement.

\section{Related work}
Several studies have employed EEG to examine various aspects of cognitive functioning and brain activity during gameplay. Aliyari et al.,2018 \cite{aliyari_alterations_2018} investigated the effects of violent and football video games on cognitive functions, cortisol levels, and brain waves, showing that playing a violent game engaged more brain regions and improved cognitive performance more effectively than playing a football game. García-Monge et al., 2020~\cite{garcia-monge_brain_2020} explored brain activity differences during various types of throwing games and found significant variations in high-beta oscillations, highlighting the influence of game types on brain activity.

Tovecký \& Siak, 2017~\cite{hostovecky_brain_2017} compared beta wave activity during 2D and 3D serious games, showing that beta waves are more active when participants are concentrating. Israsena et al., 2021~\cite{israsena_brain_2021} used a single-channel EEG headset to study the benefits of neurofeedback-based brain training games for enhancing cognitive performance in the elderly population. Their findings show significant improvements in visual memory, attention, and visual recognition. Wan et al., 2021\cite{wan_measuring_2021} investigated the impact of VR and 3D game modes on cognitive ability, discovering that VR games led to higher cognitive performance and faster reaction times.
\section{Methods}
\begin{figure*}[htbp]
    \centering
    \begin{subfigure}[t]{.24\linewidth}
        \centering
        \includegraphics[height=2.5cm]{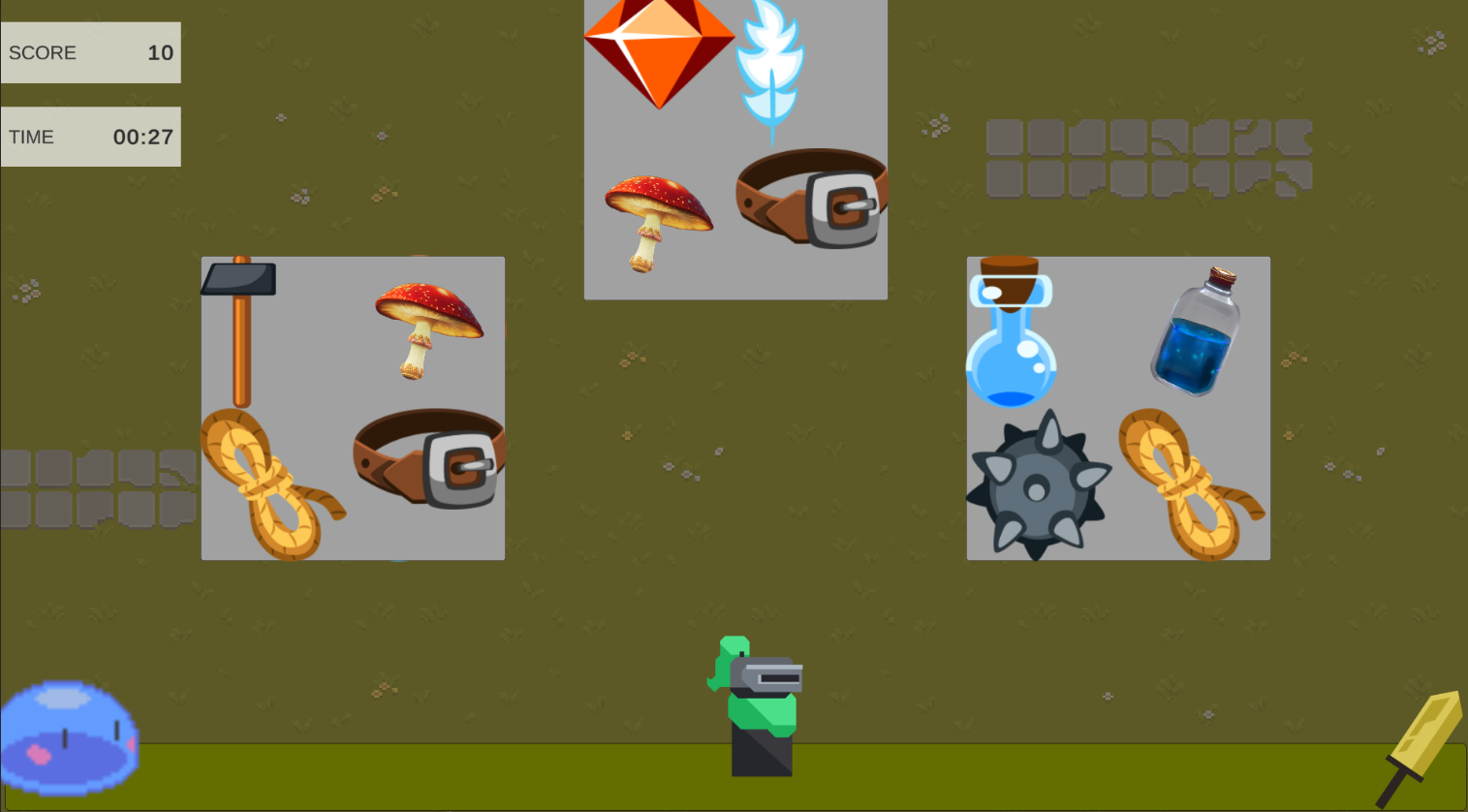}
        \includegraphics[height=2.5cm]{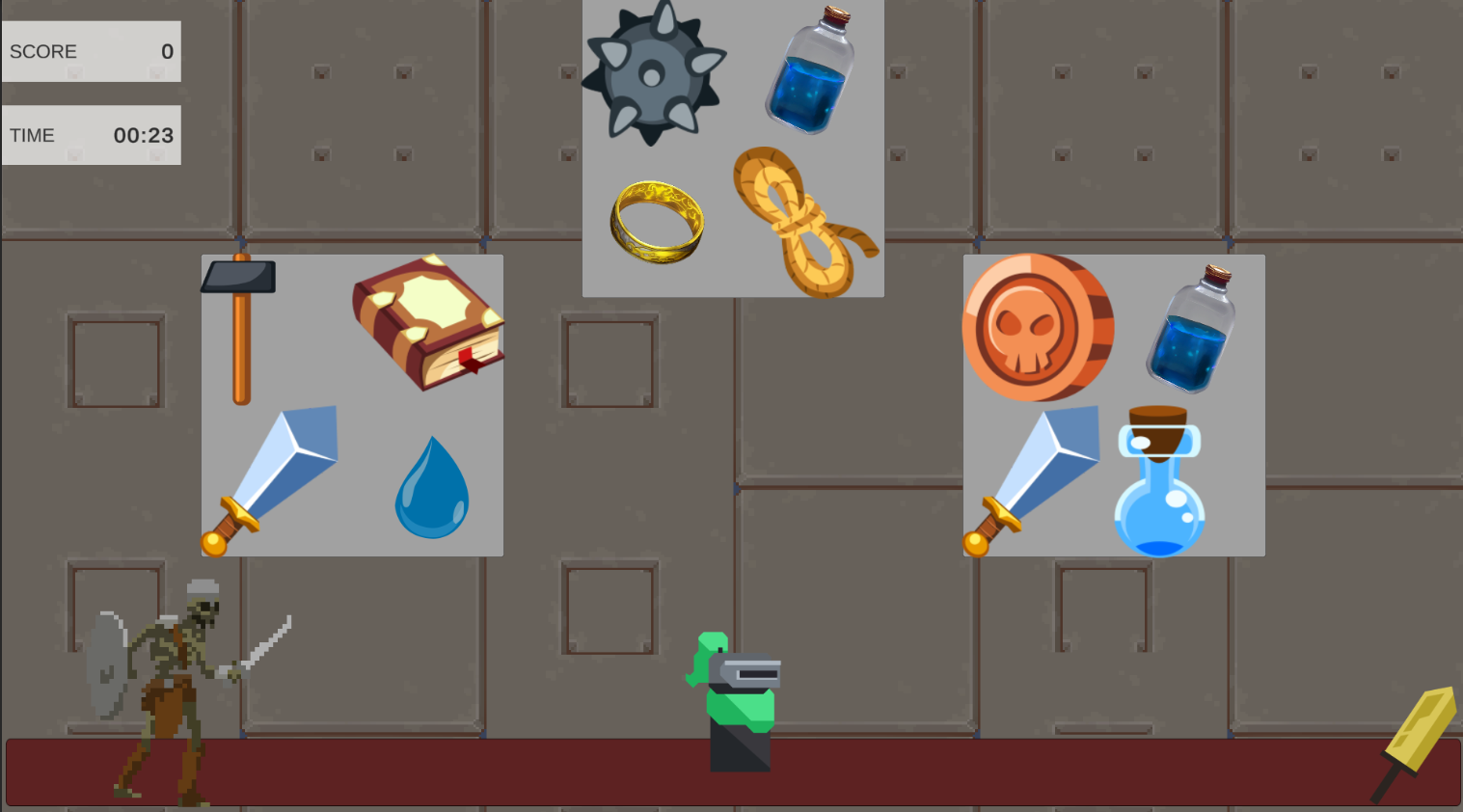}
        \caption{Dobble}
        \label{fig:dobble}
    \end{subfigure}
    \begin{subfigure}[t]{.24\linewidth}
        \centering
        \includegraphics[height=2.5cm]{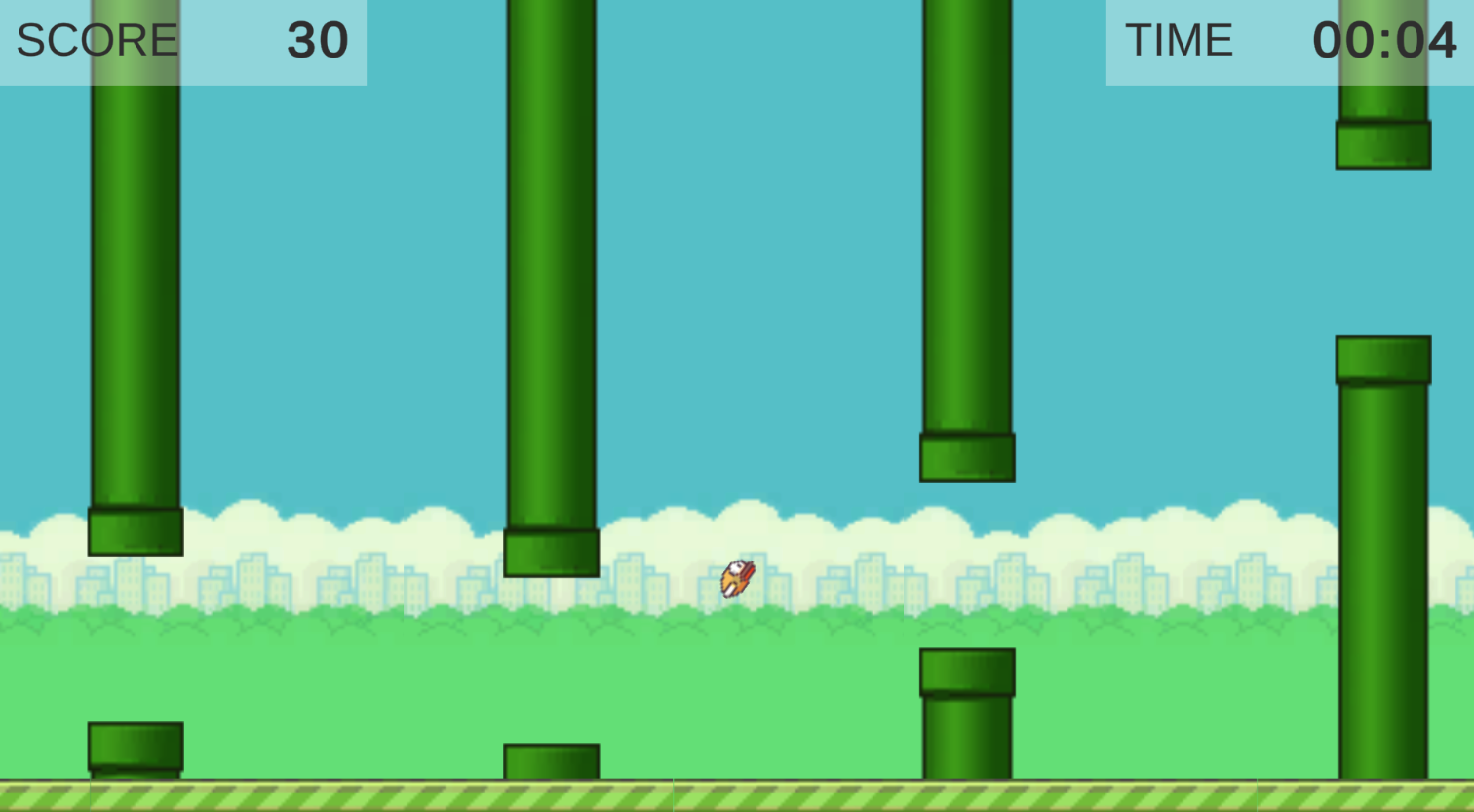}
        \includegraphics[height=2.5cm]{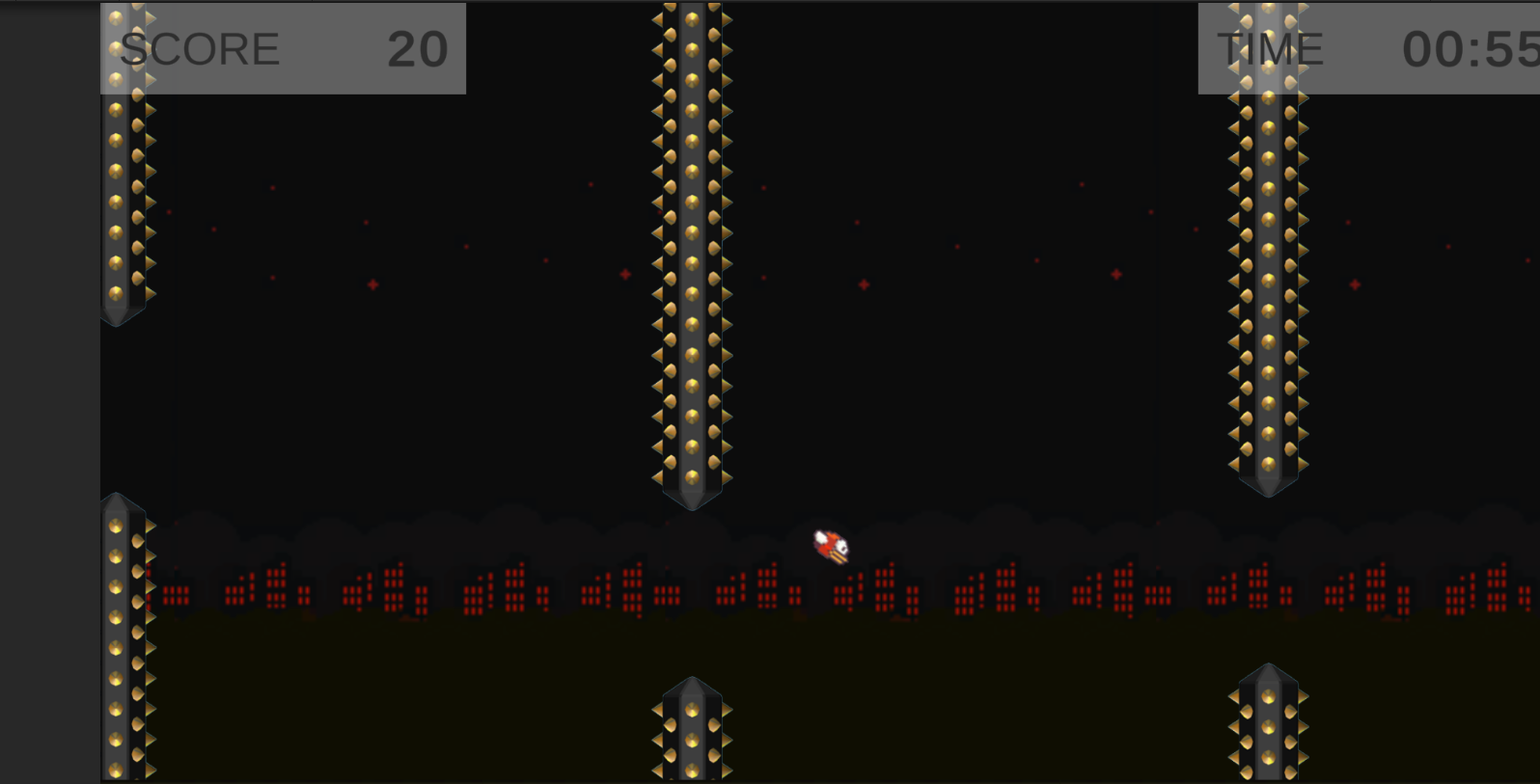}
        \caption{Flappy Bird}
        \label{fig:flappy}
    \end{subfigure}
    \begin{subfigure}[t]{.24\linewidth}
        \centering
        \includegraphics[height=2.5cm]{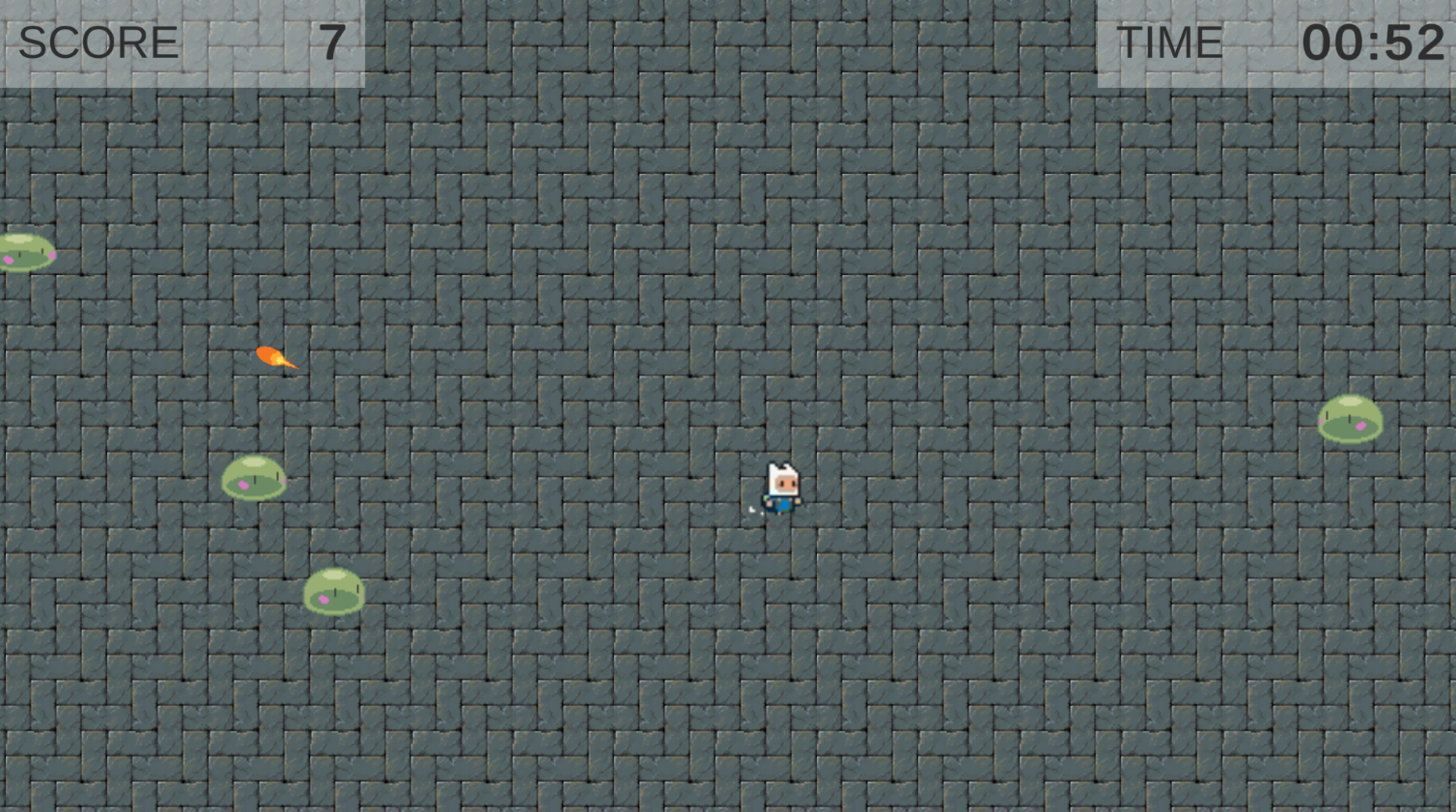}
        \includegraphics[height=2.5cm]{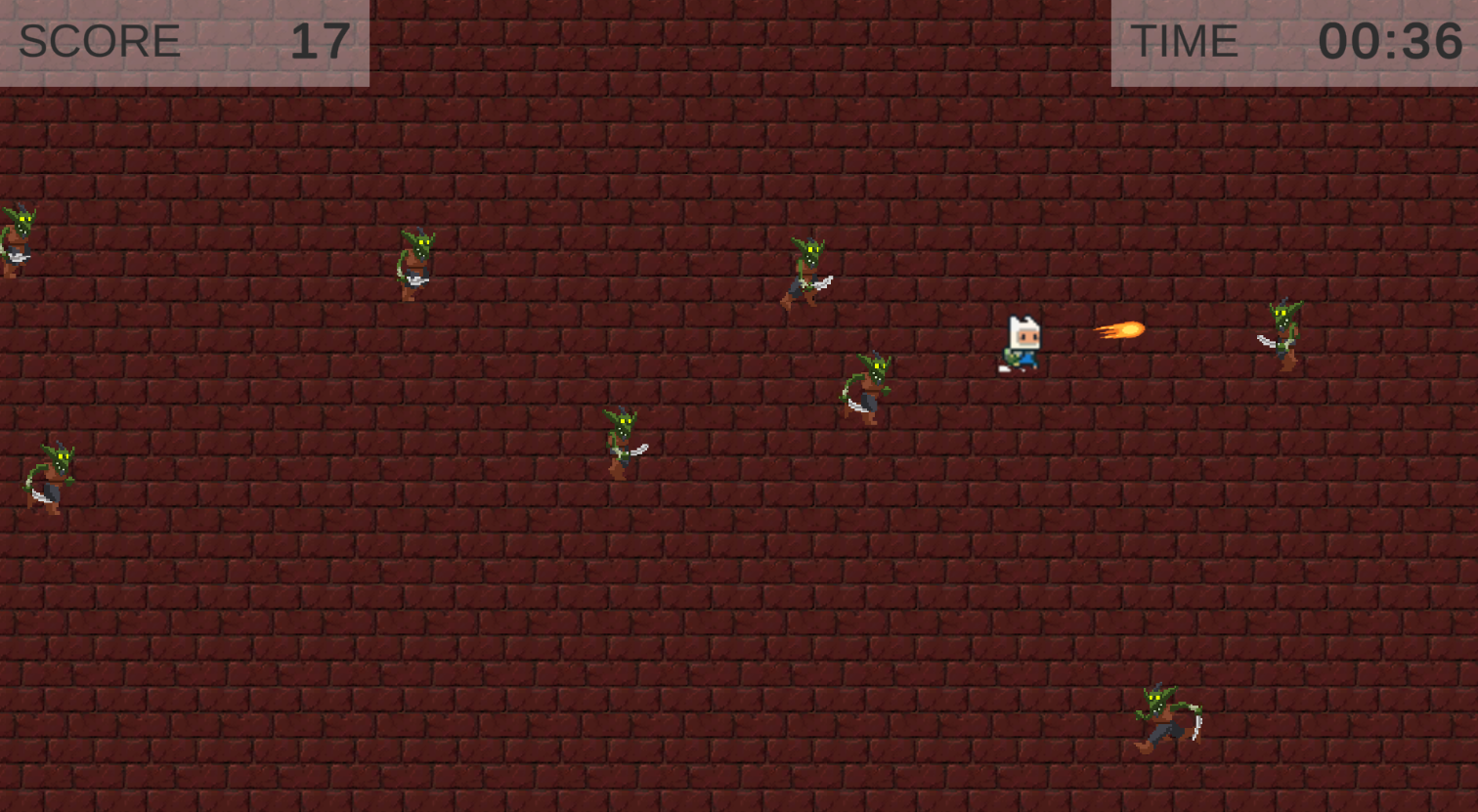}
        \caption{Shoot and Run}
        \label{fig:runandshoot}
    \end{subfigure}
        \begin{subfigure}[t]{.24\linewidth}
        \centering
        \includegraphics[height=2.5cm]{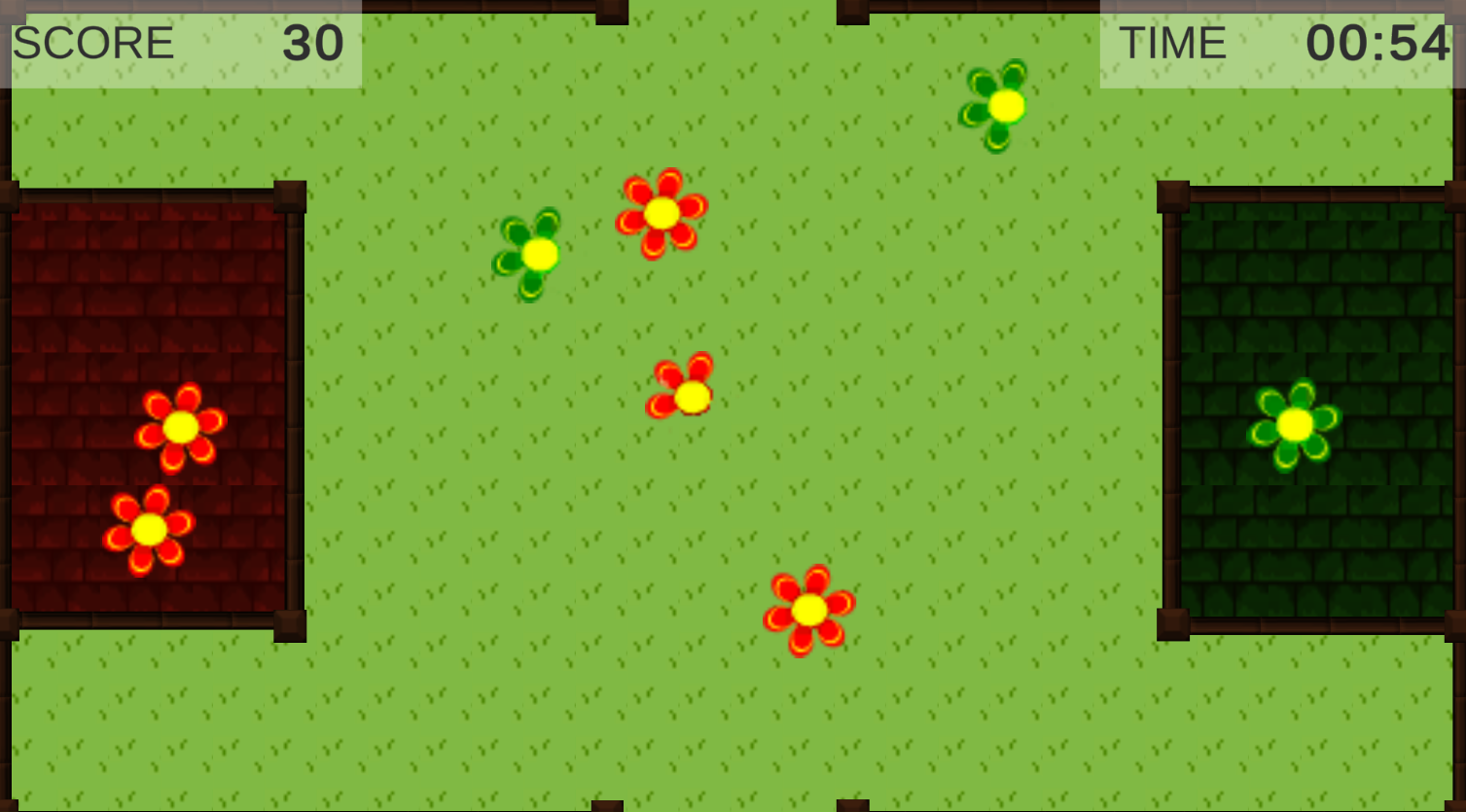}
        \includegraphics[height=2.5cm]{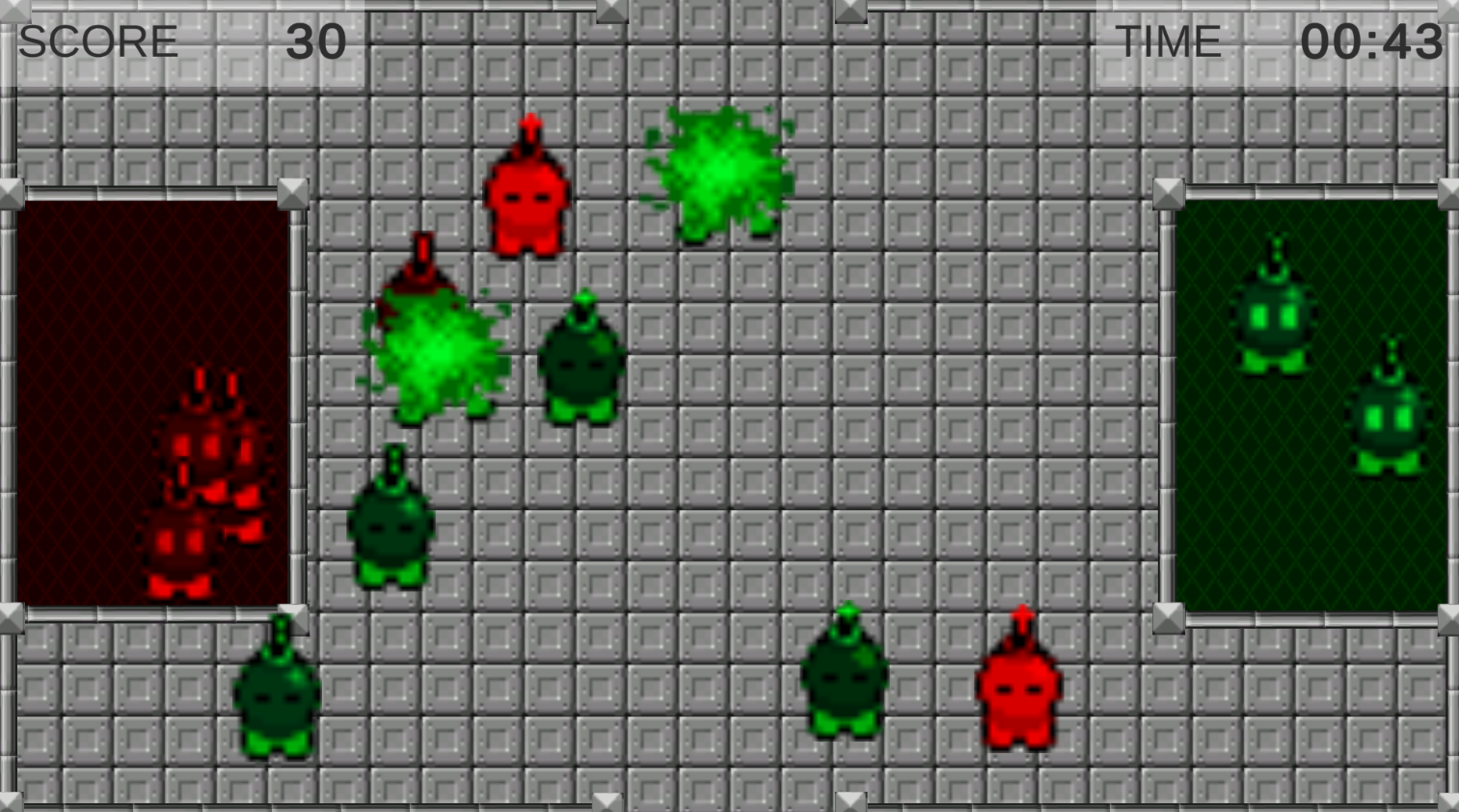}
        \caption{Sorter}
        \label{fig:sorter}
    \end{subfigure}
    \caption{Screenshots of the four games. The baseline versions are on top and the visually altered versions are below.}
    \label{fig:game_pics}
\end{figure*}
\subsection{Protocol}
12 participants were recruited, all with medium to high experience with gaming. Participants were asked to play a minigame in 60s sessions 24 times in total. We had designed 4 different minigames, with 3 versions for each game, a baseline version, a mechanically harder version and a visually altered version. Each participant is then asked to play all versions twice, in a pseudo-randomised order, to avoid repetitions of the same game. After each session, a shortened version of the CORGIS questionnaire was administered to measure perceived difficulty. During gameplay, 8 channels of EEG were recorded, and a bandpower analysis of the signal was then performed to compare game versions. EEG was recorded using Enobio 8, wet "NG Geltrodes" electrodes covering the full scalp following the 10-20 system (T7, T8, O1, O2, C1, C2, Fp1, Fp2) and sampling at 500Hz. Two ear-clipped electrodes were used as the reference. No online filter was applied. A more detailed description of each part of the setup follows below. Data and code for recreating the analysis are available at: \url{ https://github.com/itubrainlab/gamedifficultystudy}.

\subsection{Minigames and their altered versions}
Four minigames have been developed for the study, using Unity \cite{haas2014history}. Each game has three versions: 1) baseline version, 2) \textbf{mechanically harder} version with an altered core game mechanic to make the game more challenging and 3) \textbf{visually harder} version with altered graphics designed to look more threatening, but keeping the same mechanics as the baseline version. Screenshots of the games can be seen in figure \ref{fig:game_pics}.

\subsubsection{Dobble}
Inspired by the board game Dobble (fig. \ref{fig:dobble}), this game involves finding matching icon pairs on cards. As players find matches, their knight moves towards a goal while evading a pursuing monster. The knight starts in the middle, while the monster advances at a constant speed from the left. Players aim to reach the sword on the right side without getting caught. Three cards with four icons each are displayed. The top card must be dragged over a matching card, meaning it must have one common icon. The game resets if the monster catches the player or they reach the sword. In the mechanically harder version, the speed of the monster is faster, therefore requiring the player to find matches quicker to avoid losing. In the visually harder version background image and the sprite of the monster are changed. Instead of a nonthreatening-looking slime, the monster is an undead skeleton. The background is changed from a green meadow to a grey stone wall.

\subsubsection{Flappy Bird}
This minigame replicates Flappy Bird (fig. \ref{fig:flappy}, a sidescroller developed by .Gears. Players control a bird (Flappy) moving left to right, affected by gravity. Using the SPACE key, players add a vertical force to counter gravity. Losing occurs when Flappy collides with obstacles in the form of vertically shifting pairs of pipes or falls to the ground. This requires the player to use the impulse mechanic to navigate through the vertical gap between pipes. In the mechanically harder version the pipes are both vertically closer to each other and occur with higher frequency, requiring the player to both be more accurate in the jumps they make and allow for less recovery time between obstacles. In the visually hard version of the game, Flappy changed its colour from yellow to red and the background sky got reddish and darker simulating a night effect, as well as lowering the floor chroma. The obstacles design was also modified - the green pipes are replaced by spiky pillars, evoking a higher sense of danger from collisions.

\subsubsection{Shoot and Run}
Shoot and Run (\ref{fig:runandshoot} is a 2D top-down game where players control a character fighting off endless enemies. Players move using AWSD keys and shoot projectiles with the mouse, avoiding and killing enemies. Enemies spawn at a constant rate, with a low chance to spawn an extra. The player loses if an enemy reaches the player. In the mechanically harder version, enemies spawned more frequently and with an increased risk of an extra enemy spawning, overall increasing the number of enemies the player had to fight. In the visually harder version, the enemy sprites are replaced with goblins designed with acute and sharp angles, intending to make the risk of collision more salient. The movement animation was altered to be more frenetic and accentuated to make movements appear faster.

\subsubsection{Sorter}
In Sorter (fig. \ref{fig:sorter}, inspired by Nintendo's Sort or 'Splode, players sort green and red flowers into matching-coloured areas. The flowers move randomly on the screen and lose petals over time. When all petals are lost on a flower the players lose. Using the cursor, players drag continually spawning flowers to their corresponding zones, preventing petal loss. The game design encourages tracking multiple items and quick decision-making. In the mechanically harder version, the movement of the flowers was faster, requiring the player to be more accurate. In the visually harder version, the flowers are replaced by bombs threatening to explode. The state of the bombs was conveyed using increasingly redder colours for bombs closer to blowing up. Additionally, the background was changed from a green meadow to a grey metal background.

\subsection{Questionnaire}
Challenge Originating from Recent Gameplay Interaction Scale (CORGIS \cite{denisova_measuring_2020}) is a questionnaire measuring four components of the experienced challenge: performative, emotional, cognitive and decision-making challenges measured using a 5-point Likert scale. The full questionnaire includes 30 items, which is not suitable to answer 24 times after only 60s of gameplay. Additionally, not all four components are relevant to our minigames, for example, the emotional challenge covers emotionally difficult decisions. Instead, we administered a simplified version of the cognitive challenge component, including four items of originally 11 in this component. The questions asked are:
\begin{enumerate}[(a)]
    \item Playing the game requires great effort.
    \item I felt challenged when playing the game.
    \item I had to constantly keep track of what was going on in the game.
    \item I had to think actively when playing the game.
\end{enumerate}

\subsection{Bandpower preprocessing and feature extraction}
EEG signal is processed using the MNE python package \cite{gramfort_meg_2013}. A notch filter of 50 Hz is applied to remove power line noise and then a bandpass filter of [1,100] is set to remove signal drift and high-frequency noise from muscles. Subsequently, eye blink artefacts are removed using ICA ~\cite{tandle_classification_2016}. The 60s sessions are divided into 1-second segments with 100ms overlap. Then a spectral analysis is performed on each segment, using DPSS tapers to extract band power for the alpha band (8-12Hz) and gamma band (30-60Hz). The data is then normalised on the average power for the full gameplay session of that participant in each frequency band. In this study, we used only data from the two frontal electrodes Fp1 and Fp2, since frontal lobe activity is closely connected to cognitive load and decision-making \cite{collins_reasoning_2012}.

\subsection{Statistical analysis}
Both data from the questionnaire and the bandpower features are analysed using a t-test on all sessions of a game version across participants. We compare the easy version with the mechanically challenging one, and the easy version with the visually challenging one, bringing us to a total of 8 tests. A Sidak correction is applied to the significance level to allow multiple comparison testing. The only difference in analysis between the two measures is that we employ a two-tailed t-test for the questionnaire data and a one-tailed t-test for the EEG data, in accordance with our hypotheses.

\section{Results}

Results are summarised in tables \ref{questionnaire_data}, \ref{gamma_data} and \ref{alpha_data}, formatted such that the rows show the four games, and scores indicate mean value across the group. The last two columns are test results, where '-' indicate non-significance and levels of significance displayed as such: * = p$<$0.05, ** = p$<$0.005, *** = p$<$0.0005. Thresholds are shown here without correction.

\subsection{Perceived difficulty}
The results of the questionnaire analysis can be seen in table \ref{questionnaire_data}. We only obtain a single test with a significant difference between the easy version and any of the challenging versions, showing no clear evidence that perceived difficulty is consistently higher in the difficult versions in our experiment. The single test with a significant difference is the RUNSHOOT game in the mechanically more challenging version.

\subsection{Bandpower analysis}
There are two main band frequencies of interest in this study. Alpha waves, associated with lower engagement are expected to be lower in the difficult versions of the games. Conversely, gamma waves, associated with higher engagement, are expected to be higher in the difficult versions. The electrodes chosen here are the frontal electrodes (Fp1 + Fp2), as discussed earlier. 

Tables \ref{gamma_data} and \ref{alpha_data} show the results of the gamma and alpha tests respectively. The numbers reported are measured in mean bandpower across gameplay sessions of a version of a game. Power is normalised to the full gameplay session for each participant to baseline the data and make it comparable across participants. We see that all games have significantly higher gamma activity in the mechanically more challenging version, giving us a clear and consistent difference between these conditions. In the visually harder version, gamma is higher in only two of the games, showing weaker evidence for the same effect happening in this condition.

\begin{table}[htbp]
\caption{Perceived difficulty score}
\label{questionnaire_data}
\begin{tabular}{l|r|r|r|l|l}
game & baseline & mechanical & visual & test-mech & test-vis \\
\hline
DOBBLE & 14.292 & 14.958 & 13.667 & - & - \\
\hline
FLAPPY & 10.083 & 10.875 & 10.417 & - & - \\
\hline
RUNSHOOT & 12.417 & 14.667 & 11.750 & * & - \\
\hline
SORTER & 13.333 & 15.042 & 12.542 & - & - \\
\end{tabular}
\end{table}

\begin{table}[htbp]
\caption{gamma band power}
\label{gamma_data}
\begin{tabular}{l|r|r|r|l|l}
game & baseline & mechanical & visual & test-mech & test-vis \\
\hline
DOBBLE & -0.042 & 0.045 & 0.003 & *** & * \\
\hline
FLAPPY & -0.284 & -0.184 & -0.274 & *** & - \\
\hline
RUNSHOOT & -0.091 & -0.018 & -0.038 & *** & ** \\
\hline
SORTER & 0.068 & 0.128 & -0.003 & *** & - \\
\end{tabular}
\end{table}

\section{Discussion}

In all four game types, the increased mechanical challenge is associated with increased frontal lobe gamma activity. We did, however, not find any strong connection between difficulty and alpha activity, or with perceived challenge measured by the questionnaire. Instead, we found that in 2/4 of games, the visually more challenging game versions are associated with higher frontal gamma band power. 

A link between difficult gameplay and frontal gamma activity is then suggested by the results, pointing towards a possible continuous measure, that could be used for dynamic difficulty adjustment game design. This measure is reasonably lightweight, with only two electrodes used and minimal preprocessing and analysis need to extract features. 

The lack of findings in both alpha and survey data does not mean these measures are not applicable to measure difficulty necessarily. The survey is shortened from four components to one component and only includes four out of 11 items. However, the data we have seems to suggest the granularity of a 5-point scale is too low to capture the difference between the easy and hard versions of our games. Even the baseline (easy) version of the game scored on average 10-14 out of 20, which suggests participants experienced some difficulty in this version. A similar interpretation can be used to interpret the lack of down-regulation of alpha band activity in the hard versions. If the baseline versions are difficult, alpha activity might not have been very high here, since this kind of brain state is associated with drowsiness, relaxation and disengagement. If this interpretation holds, then frontal gamma activity is a promising neurological measure of challenge, able to distinguish between small differences in difficulty.

The visually difficult version, only resulted in higher frontal gamma activity in 2/4 games, suggesting that some perceived difficulty can be influenced by visuals alone. Notably, the two games with a significant effect had a clearly defined enemy which is changed to a more threatening sprite (slimes to goblins or skeletons), as opposed to obstacles. This could suggest that enemy design is especially important when deciding on a desired difficulty level. 

This work is of course limited in scope and any conclusions would need to be further tested in broader participant pools and in more varied and longer game sessions. Future work should also investigate other methods to investigate difficulty, for example using other electrode sites than frontal electrodes or utilising other frequency bands. Other candidate methods could also be to use a heart-rate monitor to measure heart rate variability or galvanic skin response, both physiological correlates of stress. 

\begin{table}[t!]
\caption{alpha band power}
\label{alpha_data}
\begin{tabular}{l|r|r|r|l|l}
game & baseline & mechanical & visual & test-mech & test-vis \\
\hline
DOBBLE & 0.104 & 0.080 & 0.046 & - & * \\
\hline
FLAPPY & -0.213 & -0.173 & -0.218 & - & - \\
\hline
RUNSHOOT & -0.156 & -0.140 & -0.134 & - & - \\
\hline
SORTER & -0.009 & -0.033 & 0.045 & - & - \\
\end{tabular}
\end{table}

\bibliographystyle{plain}
\bibliography{references}

\begin{thebibliography}{10}

\bibitem{abhang_chapter_2016}
Priyanka~A. Abhang, Bharti~W. Gawali, and Suresh~C. Mehrotra.
\newblock Chapter 2 - {Technological} {Basics} of {EEG} {Recording} and
  {Operation} of {Apparatus}.
\newblock In Priyanka~A. Abhang, Bharti~W. Gawali, and Suresh~C. Mehrotra,
  editors, {\em Introduction to {EEG}- and {Speech}-{Based} {Emotion}
  {Recognition}}, pages 19--50. Academic Press, January 2016.

\bibitem{aliyari_alterations_2018}
Hamed Aliyari, Hedayat Sahraei, Marjan Erfani, Mohamm Mohammadi, Masoomeh
  Kazemi, Mohammad~Reza Daliri, Behrouz Minaei-Bidgoli, Hassan Agaei, Mohammad
  Sahraei, Seyed Mohammad~Ali Seyed~Hosseini, Elaheh Tekieh, Maryam Salehi, and
  Fereshteh Farajdokht.
\newblock Alterations of {Cognitive} {Functions} {Following} {Violent} and
  {Football} {Video} {Games} in {Young} {Male} {Volunteers}: {By} {Studying}
  {Brain} {Waves}.
\newblock {\em Basic and Clinical Neuroscience Journal}, October 2018.

\bibitem{burelli_virtual_2013}
Paolo Burelli.
\newblock Virtual cinematography in games: investigating the impact on player
  experience.
\newblock In {\em Foundations of {Digital} {Games}: {The} 8th {International}
  {Conference} on the {Foundations} of {Digital} {Games}}. Society for the
  Advancement of the Science of Digital Games, 2013.

\bibitem{collins_reasoning_2012}
Anne Collins and Etienne Koechlin.
\newblock Reasoning, {Learning}, and {Creativity}: {Frontal} {Lobe} {Function}
  and {Human} {Decision}-{Making}.
\newblock {\em PLoS Biology}, 10(3):e1001293, March 2012.

\bibitem{csikszentmihalyi_beyond_1975}
Mihaly Csikszentmihalyi.
\newblock {\em Beyond boredom and anxiety}.
\newblock Beyond boredom and anxiety. Jossey-Bass, San Francisco, CA, US, 1975.

\bibitem{denisova_measuring_2020}
Alena Denisova, Paul Cairns, Christian Guckelsberger, and David Zendle.
\newblock Measuring perceived challenge in digital games: {Development} \&
  validation of the challenge originating from recent gameplay interaction
  scale ({CORGIS}).
\newblock {\em International Journal of Human-Computer Studies}, 137:102383,
  May 2020.

\bibitem{garcia-monge_brain_2020}
Alfonso García-Monge, Henar Rodríguez-Navarro, Gustavo González-Calvo, and
  Daniel Bores-García.
\newblock Brain {Activity} during {Different} {Throwing} {Games}: {EEG}
  {Exploratory} {Study}.
\newblock {\em International Journal of Environmental Research and Public
  Health}, 17(18):6796, September 2020.

\bibitem{gramfort_meg_2013}
Alexandre Gramfort.
\newblock {MEG} and {EEG} data analysis with {MNE}-{Python}.
\newblock {\em Frontiers in Neuroscience}, 7, 2013.

\bibitem{haas2014history}
John~K Haas.
\newblock A history of the unity game engine.
\newblock 2014.
\newblock Publisher: Worcester Polytechnic Institute.

\bibitem{hostovecky_brain_2017}
M.~Hosťovecký and B.~Babušiak.
\newblock Brain activity: beta wave analysis of {2D} and {3D} serious games
  using {EEG}.
\newblock {\em Journal of Applied Mathematics, Statistics and Informatics},
  13(2):39--53, December 2017.

\bibitem{israsena_brain_2021}
Pasin Israsena, Suwicha Jirayucharoensak, Solaphat Hemrungrojn, and Setha
  Pan-Ngum.
\newblock Brain {Exercising} {Games} {With} {Consumer}-{Grade}
  {Single}-{Channel} {Electroencephalogram} {Neurofeedback}: {Pre}-{Post}
  {Intervention} {Study}.
\newblock {\em JMIR Serious Games}, 9(2):e26872, June 2021.

\bibitem{law_systematic_2018}
Effie L.-C. Law, Florian Brühlmann, and Elisa~D. Mekler.
\newblock Systematic {Review} and {Validation} of the {Game} {Experience}
  {Questionnaire} ({GEQ}) - {Implications} for {Citation} and {Reporting}
  {Practice}.
\newblock In {\em Proceedings of the 2018 {Annual} {Symposium} on
  {Computer}-{Human} {Interaction} in {Play}}, {CHI} {PLAY} '18, pages
  257--270, New York, NY, USA, October 2018. Association for Computing
  Machinery.

\bibitem{tandle_classification_2016}
Avinash Tandle, Nandini Jog, Pancham D'cunha, and Monil Chheta.
\newblock Classification of {Artefacts} in {EEG} {Signal} {Recordings} and
  {EOG} {Artefact} {Removal} using {EOG} {Subtraction}.
\newblock In {\em Communications on {Applied} {Electronics}}, volume~4, pages
  12--19, January 2016.
\newblock ISSN: 23944714 Issue: 1 Journal Abbreviation: CAE.

\bibitem{wan_measuring_2021}
Bo~Wan, Qi~Wang, Kejia Su, Caiyu Dong, Wenjing Song, and Min Pang.
\newblock Measuring the {Impacts} of {Virtual} {Reality} {Games} on {Cognitive}
  {Ability} {Using} {EEG} {Signals} and {Game} {Performance} {Data}.
\newblock 9, 2021.

\end{thebibliography}

\end{document}